\documentclass[journal]{IEEEtran}

\ifCLASSINFOpdf
\else
\fi

\usepackage{graphicx}
\usepackage{epsfig, graphicx, amsmath, amssymb}
\usepackage{epstopdf}
\usepackage{psfrag}
\usepackage[small,compact]{titlesec}
\usepackage{wrapfig,url}
\usepackage{times}
\usepackage{multirow}
\usepackage[colorlinks, citecolor=blue]{hyperref}
\usepackage{color,soul}
\usepackage{textcomp}
\usepackage[many]{tcolorbox}
\usepackage{framed}
\usepackage{cite}
\usepackage{flushend}
\usepackage{enumitem} 
\colorlet{shadecolor}{blue!20}
\usepackage{colortbl}
\usepackage{comment,cite}
\usepackage{subfig,epsfig} 

\usepackage{stfloats}

\newcommand{\qed}{\nobreak \ifvmode \relax \else
      \ifdim\lastskip<1.5em \hskip-\lastskip
     \hskip1.5em plus0em minus0.5em \fi \nobreak
      \vrule height0.75em width0.5em depth0.25em\fi}

\newcommand{\eg}{{\it e.g., }}
\newcommand{\etal}{{\it et~al. }}
\newcommand{\ie}{{\it i.e., }}
\newcommand{\os}{{OS}}

\newcommand{\comments}[1]{}

\begin{document}

\title{The Art of CPU-Pinning:\\ Evaluating and Improving the Performance of Virtualization and Containerization Platforms}

\author{
    Davood GhatrehSamani$^+$, Chavit Denninnart$^+$, Josef Bacik$^*$, Mohsen Amini Salehi$^+$,  \\
    $^+$High Performance Cloud Computing (HPCC) lab,     University of Louisiana at Lafayette, USA\\

    \{davood.ghatrehsamani1,chavit.denninnart1,amini\}@louisiana.edu \\
    $^*$Software Engineer at Facebook Inc.\\
    josef@toxicpanda.com
}


\maketitle
\IEEEpeerreviewmaketitle
\begin{abstract}

Cloud providers offer a variety of execution platforms in form of bare-metal, VM, and containers. However, due to the pros and cons of each execution platform, choosing the appropriate platform for a specific cloud-based application has become a challenge for solution architects. The possibility to combine these platforms (\eg deploying containers within VMs) offers new capacities that makes the challenge even further complicated. However, there is a little study in the literature on the pros and cons of deploying different application types on various execution platforms. In particular, evaluation of diverse hardware configurations and different CPU provisioning methods, such as CPU pinning, have not been sufficiently studied in the literature. In this work, the performance overhead of container, VM, and bare-metal execution platforms are measured and analyzed for four categories of real-world applications, namely video processing, parallel processing (MPI), web processing, and No-SQL, respectively representing CPU intensive, parallel processing, and two IO intensive processes. 
Our analyses reveal a set of interesting and sometimes counterintuitive findings that can be used as best practices by the solution architects to efficiently deploy cloud-based applications. Here are some notable mentions: (A) Under specific circumstances, containers can impose a higher overhead than VMs; (B) Containers on top of VMs can mitigate the overhead of VMs for certain applications; (C) Containers with a large number of cores impose a lower overhead than those with a few cores. 

\end{abstract}


\section{Introduction}\label{sec:intro}

Hardware virtualization in form of virtual machines (VMs) is an indispensable part of cloud computing technology that offers isolation, manageability, consolidation, and reliability~\cite{parashar2013cloud} to cloud-based applications. However, performance overhead, resulted from several abstraction layers in the hypervisor~\cite{ZhengLi,Morabito,felter2015updated,Barik}, has historically been a side-effect of the virtualization technology. More recently, a lightweight virtualization technology, known as containerization, that provides abstraction at the application layer has gained popularity in the cloud era. Numerous cloud services, such as serverless computing~\cite{peekserverless,lloyd2018serverless} (\eg AWS Lambda~\cite{ASWLambda}, Azure Service Fabric\cite{AzureFabric}), are offered based on containers. However, we note that containerization is even making a deeper shift in application deployment, such as those used to manage critical layers of IT infrastructure. For instance, containers are being utilized in storage appliances (\eg EMC Unity~\cite{EMCUnity}) to reduce the fail-over time and improve their availability. In addition to negligible imposed overhead, containers are more storage-efficient, have shorter cloning and application deployment time, faster scale out, and offer Continuous Integration/Continuous Delivery (CI/CD)~\cite{nogueira2018improving}. Nonetheless, conflicts between processes sharing the kernel and lack of cross middleware portability are known drawbacks of the containerization~\cite{dupont2017edge}.

The pros and cons of each virtualization technology in addition to the possibility of deploying an application on bare-metal (\eg Dedicated EC2 instances in AWS) has offered cloud solution architects a range of \emph{execution platforms} (\ie bare-metal, VMs, containers, and containers on top of VMs) to deploy a certain application on the cloud. 
In particular, each application has its own characteristics that can reap the benefits of a certain type of execution platform and undergoes a different overhead. For instance, database servers are known to take advantage of bare-metal platforms, due to high volume of disk operations~\cite{zhang2012distributed}, whereas web servers benefit from \emph{virtualized platforms} (\ie VMs, containers, and containers within VMs) that offer a higher portability~\cite{yu2015live,qiu2017lxc,dupont2017edge}. 

It is challenging for a cloud solution architect to efficiently deploy a certain application via choosing a proper execution platform. In a multi-tier application, this can potentially lead to choosing distinct execution platforms for deployment of each application tier~\cite{bi2017application}. The challenge becomes further complicated when we know that the overhead of each execution platform can remarkably vary, depending on the way the execution platform is configured on the underlying hardware resources. Specifically, CPU provisioning for virtualized platforms can be configured either through CPU-quota or CPU-set (a.k.a \emph{CPU pinning})~\cite{podzimek2015analyzing}. In the former, at each scheduling event, the middleware of the host machine decides about allocating the proper CPU core(s) to each VM/container, whereas, in the latter, certain CPU cores are statically bound to each VM/container by the solution architect. Our hypothesis is that CPU pinning can drastically reduce the overhead of virtualized platforms. However, it is noteworthy that extensive CPU pinning incurs a higher cost and makes the host management more challenging. 


\begin{figure*}[b!] 
\vspace{-5px}
\centering
\subfloat[ Modules of KVM hypervisor. Each VM, called Qemu-KVM, has a full-stack of the deployed applications and an operating system. Libvirt provides necessary APIs for managing KVM.]{
{
\includegraphics[width=0.38\textwidth]{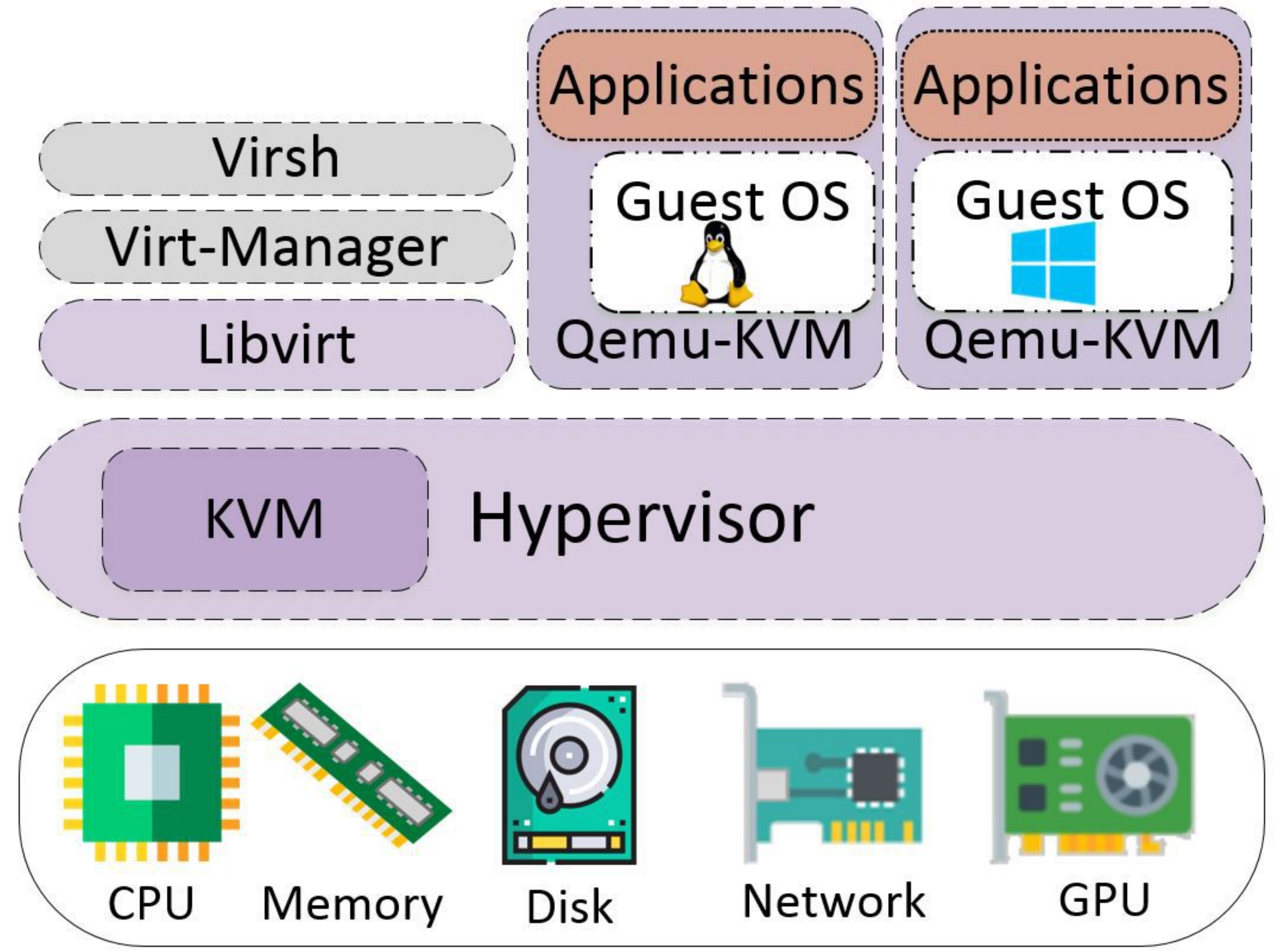}}
\label{fig:KVMarch}
}
\hspace{5pt}
\subfloat[Main modules of Docker. Containers are coupling of \texttt{namespace} and \texttt{cgroups} modules of the host OS kernel. Docker daemon interacts with Container daemon (\texttt{ContainerD}) and \texttt{runC} kernel module to manage containers.]{
{
\includegraphics[width=0.38\textwidth]{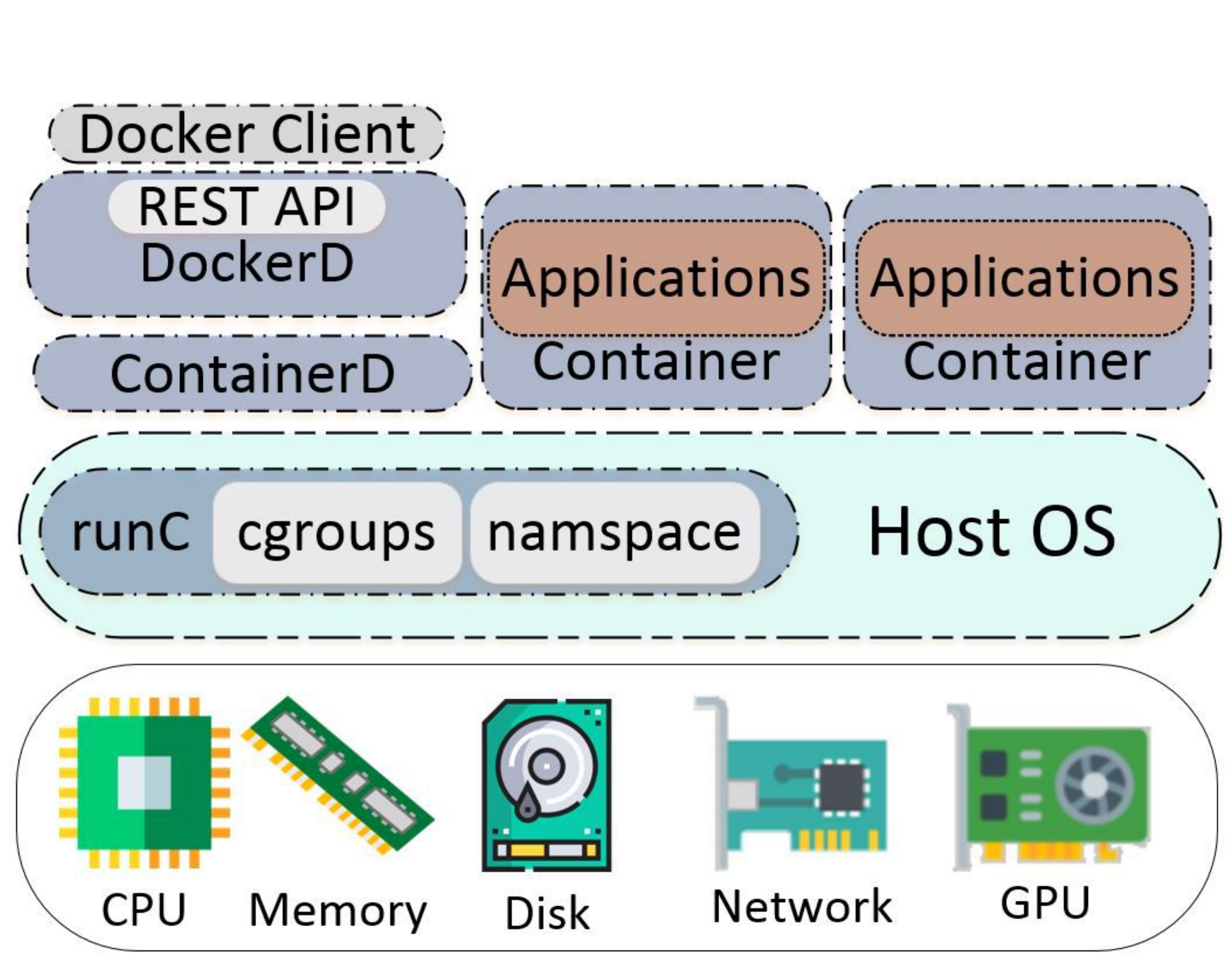} 
\label{fig:DockerArch}
}
}
\caption{High level architecture of different virtualization platforms.}
\label{fig:image2}
\end{figure*}

The \emph{goal} of this study is to unveil the imposed overhead of each virtualization platform for different application types commonly deployed on the cloud. Further, we study the impact of various CPU provisioning configurations for the underlying hardware resources. 

To achieve the goal, we conduct an extensive performance evaluation on the following four application types that exhibit different processing characteristics and are commonly used in the cloud: (A) \emph{FFmpeg}~\cite{ffmpeg16} is a video transcoding application that exhibits a CPU-bound behavior; (B) \emph{MPI}~\cite{gropp1999using} applications that represent parallel processing behavior; (C) \emph{WordPress}~\cite{patel2011performance} is a web-based system representing many short tasks with IO-bound behavior; (D) \emph{Apache Cassandra}~\cite{abramova2013nosql} is a NoSQL database management system representing an extensive IO-bound behavior within a single large process.

Each application type is deployed under various resource configurations (a.k.a instance type) on our private cloud. Specifically, we evaluate the imposed overhead of different execution platforms using different number of CPU cores and under two circumstances---when CPU pinning is in place and when it is not. In summary, the \emph{contributions} of this paper are as follows:
\begin{itemize}
  \item Measuring and analyzing the imposed overhead of different execution platforms for widely-used cloud-based applications.
  \item Analyzing the impact of altering resource configurations, including number of CPU cores and CPU pinning, on the imposed overhead of different execution platforms.
  \item Proposing a set of best practices for deploying different application types in different virtualization platforms.
\end{itemize}


This paper is structured as follows. Section \ref{sec:background} provides an architectural view of virtualization platforms and describes CPU provisioning models. Section \ref{sec:PerfStudyanalysis} describes the applications, our testbed, and analyzes the overhead of different platforms under various configurations. In Section \ref{sec:cross}, we perform a cross-application overhead analysis of the imposed overhead across different application types. Section \ref{sec:relwk} presents the most relevant works in the literature for benchmarking virtualization platforms and the use of pinning. Then, in Section \ref{sec:conclsn}, we summarize the lessons learnt and provide a set of best practices to efficiently configure cloud platforms.

\section{Background}
\label{sec:background}
\subsection{Overview}
Virtualization platforms emulate and isolate compute, storage, and network resources within a host. Current virtualization platforms are categorized based on the level of abstraction they provide. In particular, VMs provide a hardware layer abstraction, known as \emph{hardware virtualization}, whereas containers enable abstraction from the operating system (OS) layer, known as \emph{OS virtualization} \cite{BuyyaMasteringBook}.


\subsection{Hardware Virtualization (VM)} 
Hardware virtualization operates based on a hypervisor that enables one or more isolated guest operating systems (VMs) on a physical host \cite{ppACSC,salehi2011contention}. KVM \cite{felter2015updated} is a popular open-source hypervisor extensively used by cloud providers. For instance, AWS developed a KVM-based hypervisor called \emph{Nitro} \cite{awsnitro} and uses it for its \texttt{C5} VM types. Many datacenter management platforms have been developed around KVM hypervisor. For instance, Hyper Converged Infrastructure (HCI) platforms \cite{Hci} (\eg Nutanix \cite{Nutanix}, Maxta \cite{Maxta}, and Cloudistics \cite{Cloudistics}) that enable integrated software defined datacenters have adopted KVM as their underlying hypervisor.

KVM hypervisor, depicted in Figure~\ref{fig:KVMarch}, is part of the host kernel and works directly with hardware virtualization features (\eg \texttt{Intel-VT} and \texttt{AMD-v} \cite{goto2011kernel}) to share the hardware resources across multiple fully isolated VMs. KVM interfaces with the user space (\eg via \texttt{QEMU}) and executes guest OS commands via \texttt{ioctls} kernel module. Qemu-KVM emulates hardware devices for VMs and can para-virtualize \cite{goto2011kernel} IO devices, such as disk and NIC, by using specific drivers that expedite accessing them. \texttt{Libvirt} is a library to help external services to interact with KVM/QEMU using its APIs. \texttt{virsh} and \texttt{virt-manager} are command-line and GUI interfaces for Libvirt. 

\subsection{OS Virtualization (Container)} 
Container is a lightweight and portable virtualization solution in which the host OS kernel is shared across multiple isolated user-space environments. In contrast to VMs, containers are transparent from the host OS perspective. That is, the processes created by a container are visible to the host OS as native processes, however, the container itself is not a process. All processes created via a container have access to the same set of resources and libraries. 

Technically speaking, a container is an abstraction created by the coupling of \emph{namespace} and \emph{cgroups} modules of the host OS. A namespace that is assigned to a container creates an abstraction for the processes of the container and offers them an isolated user space, such as network configurations, storage space, and software packages. 

Control Groups (a.k.a \texttt{cgroups}) is a kernel module that enforces and monitors resource usage limitations for a given group of processes \cite{Kernel.org-cgroup}\cite{Josef}. Cgroups module of the host OS is in charge of enforcing resource constraints designated for a container instance. As an example, for a container instance with two CPU cores, cgroups oversees its usage not to go beyond the two cores across the entire host. It is noteworthy that the way cgroups enforces constraints is a decisive factor from the performance overhead perspective.

Docker \cite{yu2015live} is the most widely adopted container technology in the cloud era, hence, we consider it as the container platform in this study. However, we believe that our findings can be extrapolated to other containerization techniques that operates based on cgroups (\eg Singularity). Figure~\ref{fig:DockerArch} illustrates the architecture of the Docker container. Docker Engine (\texttt{DockerD}) receives container management requests via its APIs. The engine is in charge of creating container instances and enforcing their specifications via \texttt{containerD} service. 
ContainerD utilizes \texttt{runC} module of the OS kernel to create namespace and cgroups for each container instance.

\subsection{CPU Provisioning for Virtualized Platforms}\label{subsec:virt}
The host OS generally uses time sharing scheduling policies (\eg Completely Fair Scheduler (CFS) \cite{wong2008fairness}) that does not specify a processing unit (\eg core) to each process. That is, a VM or a container-based process is assigned to a different set of CPU cores of the host machine in each quantum, during their life cycle. In this study, a platform that has its provisioned CPU cores in this default manner is called to be deployed in the \emph{vanilla} mode. 

Alternatively, a user can choose the \emph{pinning} mode, to manually set the CPU cores allocated to a given process (either a VM or a container) and override the default host OS scheduler. We explained the technical details of how to perform CPU pinning in a web-blog post \footnote{How to perform CPU pinning: \url{https://bit.ly/2XrENUM}}. In this case, the host OS scheduler allocates the pinned process only to the specified cores. Note that, unlike vanilla mode that utilizes all the host CPU cores to cumulatively offer the expected performance, for a pinned platform, only the designated cores are utilized and the rest are left idle. As such, the side-effect of pinning (\ie not using the host \os~scheduling) can appear in a lower CPU utilization, hence, it should not be used carelessly for any application.

\section{Overhead Analysis of Different Application Types on a Variety of Virtualized Platforms}
\label{sec:PerfStudyanalysis}
\subsection{Evaluation Environment}
In this section, we evaluate the performance of four cloud-based application types (detailed in Table \ref{table:apptypes}) on four popular execution platforms.

\begin{table}[h]
\centering
    \vspace{-2px}
\begin{tabular}{ c|c|c } 
 \hline
 \textbf{Type} &  \textbf{Version} & \textbf{Characteristic} \\
 \hline
 \hline
 FFmpeg & 3.4.6 & CPU-bound workload\\
 Open MPI & 2.1.1 & HPC workload \\ 
 WordPress & 5.3.2 & IO-bound web-based workload \\ 
 Cassandra & 2.2 & Big Data (NoSQL) workload \\ 
  \hline
\end{tabular}
\caption{\small{Specifications of application types used for evaluation.} }
\label{table:apptypes}
   \vspace{-5px}
\end{table}

The performance metric we measure in the evaluations is the \emph{execution time} of each application type. Also, to quantify the overhead of a certain virtualized platform, we define \emph{overhead ratio} as the average execution time offered by a given virtualized platform to the average execution time of bare-metal. 
Table \ref{table:hwconfig} describes the configuration of instance types used for the evaluation. The host server is a DELL PowerEdge R830 with 4$\times$Intel Xeon E5-4628Lv4 processors with 112 homogeneous cores, 384 GB memory (24$\times$16 GB DRAM), and RAID1 (2$\times$900 GB HDD) storage. Each processor is 1.80 GHz with 35 MB cache and 14 processing cores (28 threads).

\begin{table}[h]

\centering
\begin{tabular}{ c|c|c } 
 \hline
\textbf{Instance Type} &  \textbf{No. of Cores} & \textbf{Memory (GB)} \\
 \hline\hline
 \texttt{Large} & 2 & 8 \\
 \texttt{$\times$Large} & 4 & 16 \\ 
 \texttt{2$\times$Large} & 8 & 32 \\ 
 \texttt{4$\times$Large} & 16 & 64 \\ 
 \texttt{8$\times$Large} & 32 & 128\\
 \texttt{16$\times$Large} & 64 & 256\\
  \hline
\end{tabular}
\caption{\small{List of instance types used for evaluation.}}
\label{table:hwconfig}
   \vspace{-5px}
\end{table}

\begin{figure*}[t]
  \centering
  \includegraphics[width=0.8\textwidth]{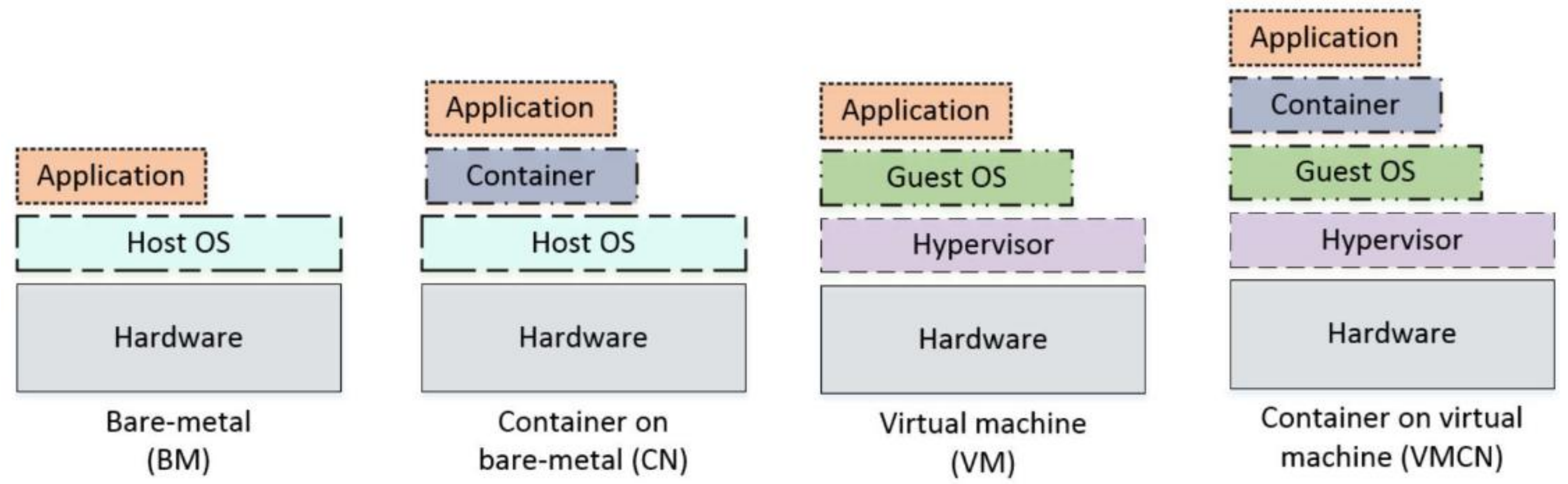}
  \caption{The four execution platforms used for performance evaluation of different application types.}
  \label{Fig:Platforms}
  \vspace{-0.2in}
\end{figure*}

The four studied execution platforms include bare-metal, which imposes the minimum overhead and is used as the baseline, in addition to three variations of virtualized platforms commonly used in the cloud (\ie VMs, containers, and containers within VMs). Figure \ref{Fig:Platforms} provides a schematic view of the four execution platforms and Table~\ref{table:config} elaborate on the specifications of each platform. The abbreviations mentioned in the table are used henceforth to represent each execution platforms. Note that each execution platform can be instantiated using any instance type of Tabel~\ref{table:hwconfig}. 

\begin{table}[hb]
\vspace{-2px}
\centering
\begin{tabular}{ c|c|l } 
 \textbf{Abbr.} & \textbf{Platform} &  \textbf{Specifications} \\
 \hline
 \hline
 BM & Bare-Metal & \small{Ubuntu 18.04.3, Kernel 5.4.5} \\
 \hline
VM & \small{Virtual Machine} & \small{Qemu 2.11.1, Libvirt 4}\\
& &\small{Ubuntu 18.04.3, Kernel 5.4.5} \\
\hline
 CN & \small{Container on} & \small{Docker 19.03.6}, \\
 & \small{Bare-Metal} & \small{Ubuntu 18.04 image} \\ 
 \hline
VMCN & \small{Container on VM} & \small{As above}\\
\end{tabular}
\caption{\small{Characteristics of different execution platforms used in the evaluations. First column shows the abbreviation of the execution platform used henceforth in the paper.}}
\label{table:config}
\end{table}

Bare-metal (BM) execution platform only includes the host OS and the application. In VM platform, one Ubuntu VM instance is created (based on KVM hypervisor) to process requests of the application type in question. 
Similarly, in container platform (CN), one Docker container is instantiated on bare-metal from an Ubuntu image. Lastly, VMCN platform refers to an execution platform where a Docker container is instantiated within a VM (with the aforementioned configurations).

Resource contention between coexisting processes in a host can potentially affect the tasks' execution times, hence, introducing noise in our overhead measurement objective. To avoid such noises, we assure that 
each application type is examined in isolation. That is, during the evaluation process, there is no other coexisting workload in the system.

As for the performance monitoring tools employed in this study, we used basic Linux utilities, such as \texttt{top, htop, iostat} and \texttt{perf}, alongside with \texttt{BCC} (BPF Compiler Collection \cite{BCC}) as a profiling tool to perform kernel tracing and to track the execution status of processes running inside the OS. In particular, we used \texttt{cpudist} and \texttt{offcputime} to monitor and profile the instantaneous status of the processes in the OS scheduler.

Irrespective of the execution platform, the host OS scheduler is the ultimate decision maker in allocating processes to CPU cores \cite{wong2008fairness}. A process (\eg \texttt{FFmpeg}) can be potentially assigned to a different set of cores at each scheduling event. It is important to note that even VMs are considered as processes from the host OS perspective. 

As the scheduling events happen frequently and there are numerous cores in a typical cloud host, migrating processes from one core to another at each event is expected to induce a remarkable overhead in the cloud hosts. Specifically, among other reasons, migrating a given process induces overheads for redundant memory access due to cache miss, reestablishing interrupts for IO operation, and context switching \cite{wong2008fairness,qiu2017lxc}. Even more overheads are involved in migrating virtualized platforms, \eg for resource usage accounting activities.

We need to measure and verify the significance of the induced overhead of different execution platforms on the overall performance. We envisage that overriding the host OS scheduler, via \emph{CPU pinning} techniques, limits process migrations to a smaller set of CPU cores, hence, reduces the imposed overhead. As such, to verify the impact of CPU pinning, we evaluate each execution platform for different workloads with and without CPU pinning. Note that the virtualized platforms offer built-in pinning ability (\eg via $Qemu$ configuration file for each VM). For BM, we modelled pinning via limiting the number of available CPU cores on the host using GRUB configuration in Linux \cite{marshall2001boot}.


\subsection{Application-Specific Overhead Analysis}

In the following subsections, we concentrate on the characteristics of each application type across different execution platforms. Later, in Section~\ref{sec:cross}, we provide a collective cross-application overhead analysis on the root causes of the imposed overhead. 

\subsubsection{Video Processing Workload Using FFmpeg}\label{subsec:ffmpeg}
FFmpeg offers a wide variety of video transcoding functions, such as those to change video resolution, bit-rate, frame rate, and compression standard. Changing the compression standard (a.k.a \emph{codec}) is known to be the most CPU-intensive transcoding operation \cite{CVSSJournal,analysistpds18} with small memory footpring (around 50 MB in our observations). Hence, we employ it in this study to represent a CPU-intensive workload. This also makes the overhead of the execution platform more visible and makes it easier to harvest. FFmpeg is a multi-threaded application and can utilize up to 16  CPU cores to transcode a video. Hence, for this evaluation, we do not allocate more than 16 cores (\ie \texttt{4$\times$large}) to each execution platform. 

We examine a source video segment\footnote{The video file is free-licensed and is publicly available in the following address:~\url{https://peach.blender.org/download/}} that has a large codec transcoding time. The reason that we examine one video segment is to concentrate on the overhead resulted from the execution platform and remove any uncertainty in the analysis, caused by the video characteristics. 
The source video segment is 30 MB in High Definition (HD) format. The codec is changed from AVC (\texttt{H.264}) to HEVC (\texttt{H.265}). The evaluation was conducted 20 times and the mean and confidence interval of the results were collected. 

\begin{figure}[ht]
  \centering
  \includegraphics[width=0.45\textwidth]{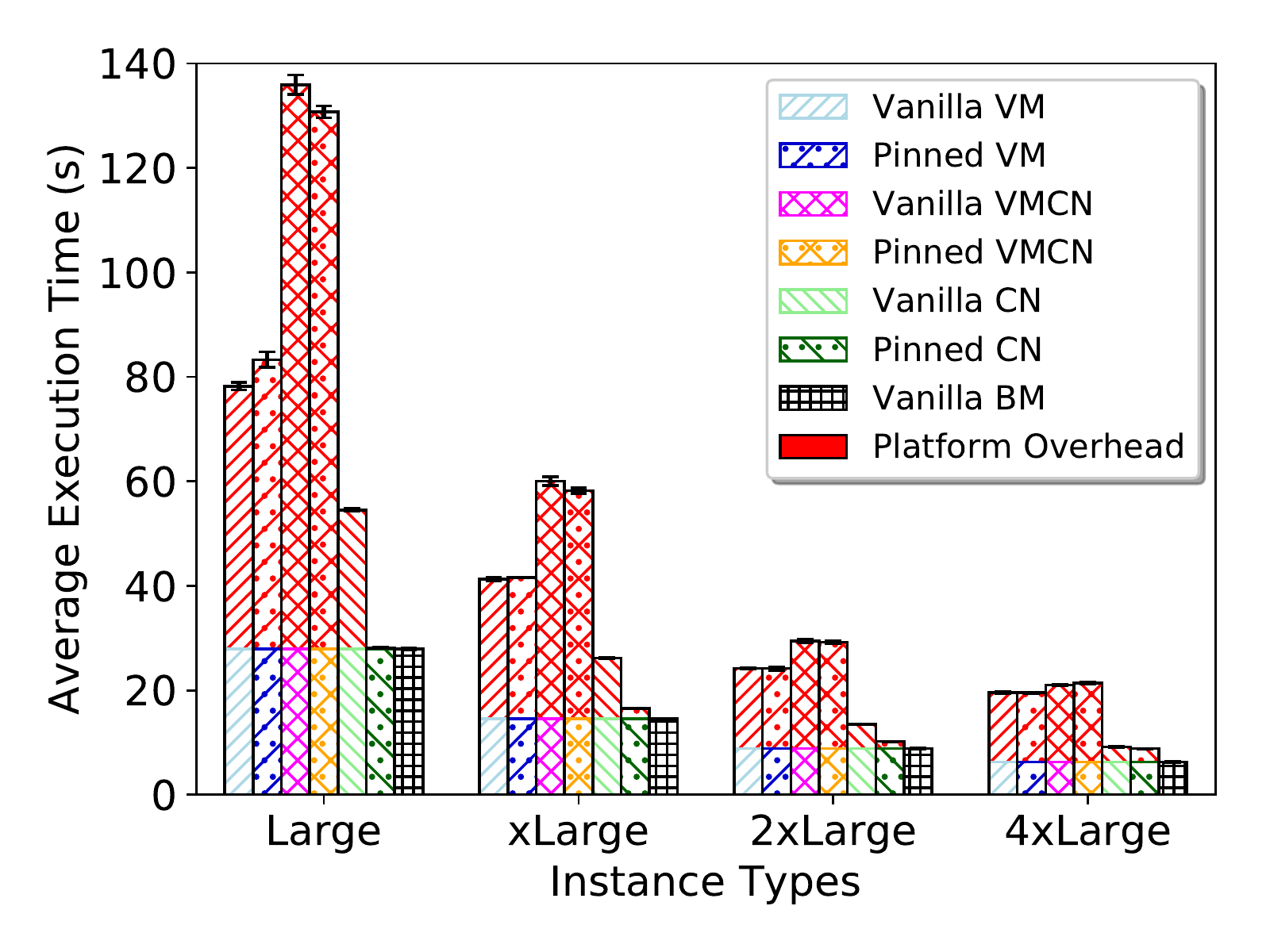}
  \caption{\small{Comparing execution time of FFmpeg on different execution platforms under varying number of CPU cores. Horizontal axis indicates the number of CPU cores in form of different instance types.}}
  \label{Fig:FFmpeg-all}
   \vspace{-0.1in}
\end{figure}

Results of the evaluation is shown in Figure~\ref{Fig:FFmpeg-all} where the vertical axis shows the mean execution time for each experiment and the horizontal axis shows different instance types. We note that the confidence interval in many cases were negligible. 

Specific observations and analysis of Figure~\ref{Fig:FFmpeg-all} are enumerated in the following list. Note that, we defer further analysis of these observations to Section~\ref{sec:cross} where we conduct a comparative study across all application types. 

\begin{enumerate}[label=\roman*.]
\item VMCN imposes the highest overhead and pinning it cannot reduce the overhead remarkably. Alternatively, CN platforms (particularly, pinned CN) are shown to impose the minimal overhead with respect to BM. Importantly, we observe that as the number of cores increases, the overhead of vanilla CN and both VMCN platforms decrease. 

\item The imposed overhead of VM platforms (vanilla or pinned) across all instance types is remarkable to the extent that causing the execution times to remains at least twice as much as BM. Unexpectedly, pinning does not mitigate the imposed overhead for VMs when FFmpeg application is deployed.

\item By adding the containerization layer on top of VM (\ie VMCN), even a larger performance overhead is imposed. The maximum and minimum imposed overhead ratios are 4 and 1, respectively. However, as the number of CPU cores allocated to the VMCN increases, the overhead is mitigated drastically, such that for \texttt{4$\times$Large}, the overhead imposed by VMCN is almost the same as VM. Furthermore, overhead ratio of VM and VMCN (vanilla and pinned) remains almost the same across all instance types.

\item Unlike VM and VMCN, pinning CN significantly reduces the overhead, particularly when containers are allocated with fewer processing cores (\eg \texttt{Large}). This suggests that pinned CN is a suitable virtualization platform for CPU-bound applications, such as FFmpeg. 

\end{enumerate}
   
\subsubsection{Parallel Processing Workload Using MPI} 
Message Passing interface (MPI) \cite{gropp1999using} is a widely-used high performance computing (HPC) platform to develop parallel programs. For the analysis, we examined two MPI applications, namely \emph{MPI\_Search} (for parallel searching of an integer value) \cite{FSU} and \emph{Prime\_MPI} \cite{FSU2} (to find all prime numbers within a given range). In these applications, the communication part dominates the computation part. This is to enable us concentrating on the impact of various virtualization platforms on the overall overhead imposed, in circumstances where intensive communication occurs between cores of the same virtualized platform. As our observations for both of the MPI applications were alike, to avoid redundancy, we only report the results for MPI\_Search. To remove any randomness in the results, the evaluations were conducted 20 times and the mean and confidence interval of the execution times are reported. 

\begin{figure}[t]
  \centering
  \includegraphics[width=0.45\textwidth]{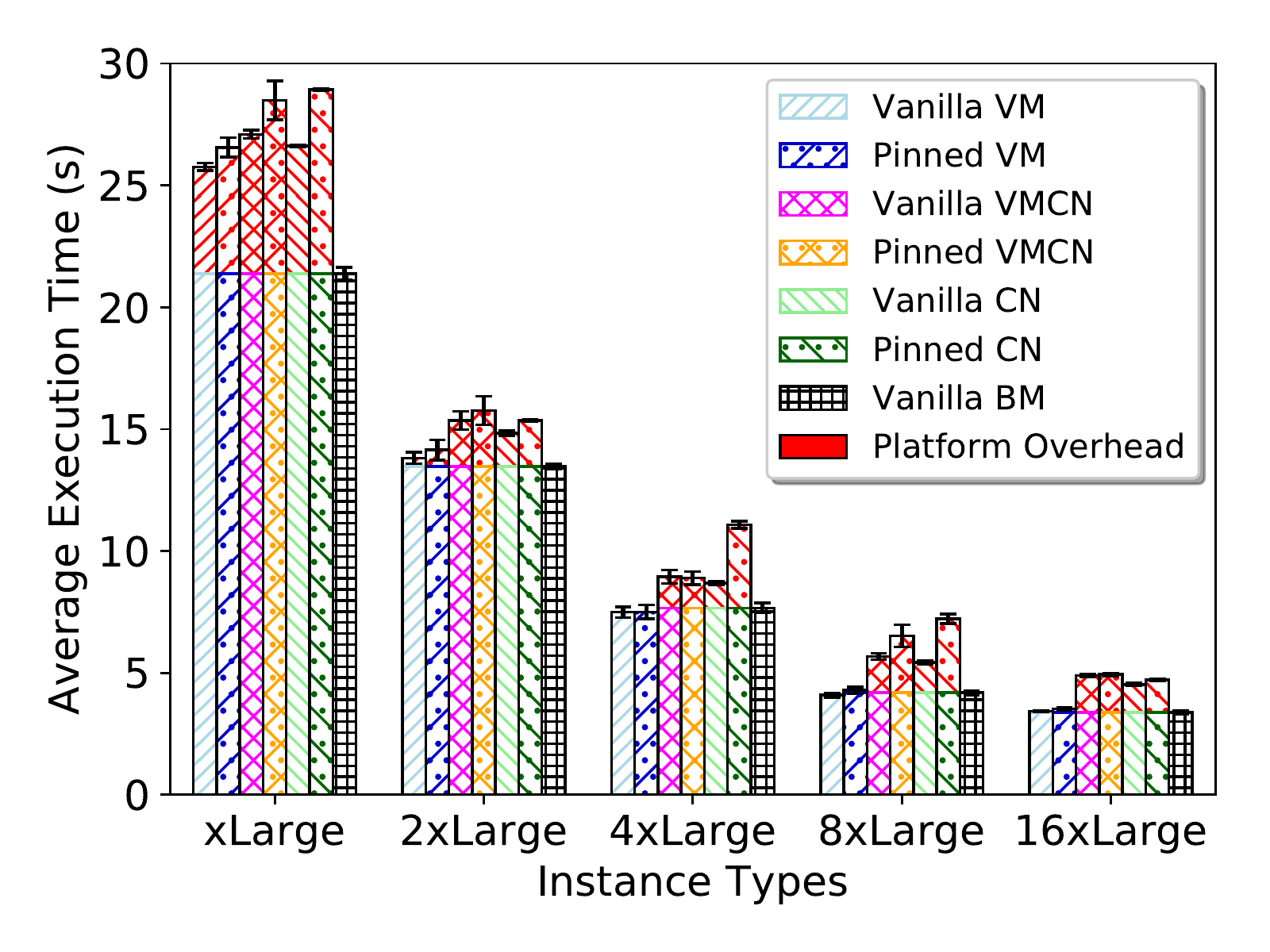}
  \caption{\small{Comparing execution time of MPI\textunderscore search on different execution platforms. Horizontal axis represents the number of CPU cores in the form of different instance types. Vertical axis shows the mean execution time (in seconds).}}
  \label{Fig:MPI-all}
   \vspace{-5px}
\end{figure}

Result of this evaluation is shown in Figure \ref{Fig:MPI-all}. 
Our observations and analyses are as follows:
\begin{enumerate}[label=\roman*.]
\item Overhead imposed by VMCN platforms (both vanilla and pinned) is slightly more than VM-based platforms. Surprisingly, the overhead of CN (vanilla and pinned) even exceeds the VMCN platforms. A closer observation reveals that, for the MPI applications, the overhead of any containerized platform exceeds the ones for the VM platforms.
Although the difference in the imposed overheads is reduced for larger number of cores, the overhead ratio remains the same.

\item From \texttt{2$\times$Large} onward, we notice that the overhead of VM platforms (vanilla and pinned) becomes negligible and the execution times become almost the same as BM. The reason is that, as we move towards larger instances, the bottleneck shifts from computation part to the communication part. Because the hypervisor (KVM) provides an abstraction layer to facilitate inter-core communication between VM's cores, the imposed overhead diminishes and their execution times approach BM. This is unlike communications within cores of a container that involves host OS intervention, thus, implies a higher overhead \cite{ZhengLi}. It is noteworthy that for smaller instance types, the bottleneck of VM platforms is in the computation part that, due to several abstraction layers, incurs a significant overhead.

\end{enumerate}

\subsubsection{Web-based Workload Using WordPress}
WordPress is a PHP-based content management system (CMS) that uses Apache Web Server and MySQL database in the back-end. It is broadly deployed in clouds 
and known to be IO intensive \cite{patel2011performance}, because each HTTP request to a website implies at least network (to read/write from/to the socket) and disk IO (to perform file/database operations). As such, we consider it as a representation of an IO intensive application, in which each process receives at least three IO interrupts. 

We configured the same WordPress website on all of the execution platforms. Apache Jmeter \cite{Jmeter} is a tool to apply workload and measure the performance of a web server. We configured it to generate 1,000 simultaneous web requests (\ie processes) on each execution platform and then, we calculated the mean execution time (a.k.a \emph{response time}) of these web processes. We note that Jmeter itself is a resource intensive application that can affect our intended performance evaluation. Therefore, we configured it to run on a dedicated server in the same network. To remove any possible environmental randomness, we evaluated the workload six times on each instance type. Then, we report the mean and 95\% confidence interval of response time of all web requests.

\begin{figure}[b]
 \vspace{-7px}
  \centering
  \includegraphics[width=0.45\textwidth]{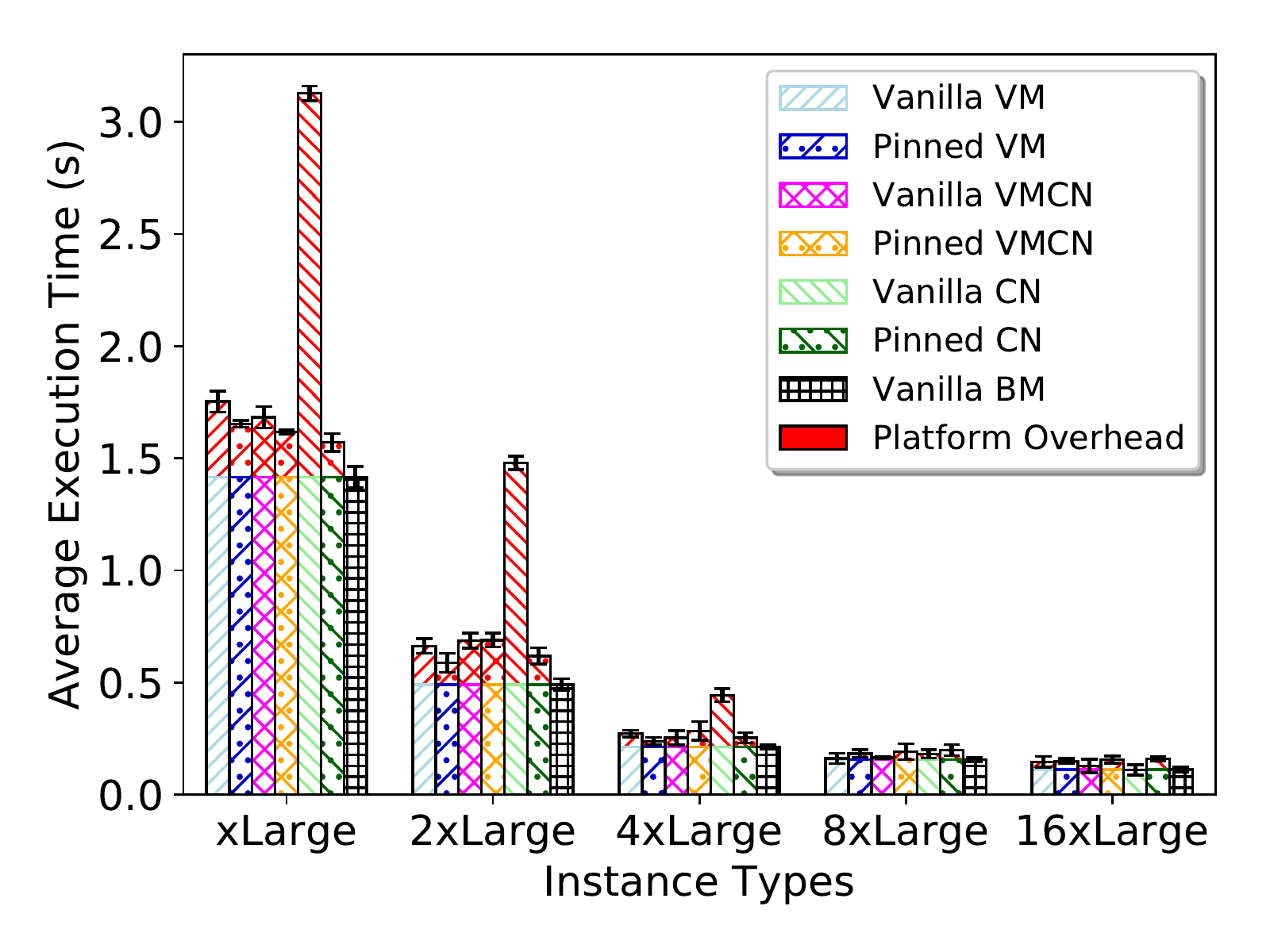}
  \caption{\small{Comparing mean response time (aka execution time) of 1,000 web processes on different execution platforms (WordPress evaluation). The horizontal axis represents the number of CPU cores in the form of different instance types and the vertical axis shows the mean execution time (in seconds).}}
  \label{Fig:Wordpress-all}

\end{figure}

Results of this evaluation are shown in Figure \ref{Fig:Wordpress-all}. The vertical axis shows the mean execution time of 1,000 web processes (in Seconds) and the horizontal axis shows different instance types. Our observations and analyses of the results are enumerated in the following list: 

\begin{enumerate}[label=\roman*.]
\item Unlike pinned CN that imposes the lowest overhead, vanilla CN imposes the highest overhead across all execution platforms---twice as much as BM for \texttt{Large} instance type. However, by increasing the number of CPU cores, this behavior is changed such that the mean execution time offered by vanilla CN approaches BM. As a similar behavior is observed for other application types, we defer analysis of this observation to Section~\ref{sec:cross}.

\item Even though VMCN platforms (vanilla and pinned) include one more layer of virtualization in compare with VM platforms, they impose a slightly lower overhead. Further analysis within VM platforms shows that the pinned VM consistently imposes a lower overhead than the vanilla VM. Building upon these two observation, we hypothesize that, for IO intensive applications, both pinning and containerization are decisive factors in mitigating the imposed overhead of the virtualized platform. In curtail, the reason that pinning remarkably mitigates the execution time overhead is a more efficient use of the cache and the possibility to pin virtualized platforms on CPU slots based on their IO affinity \cite{pinning}. Alternatively, in a non-pinned (\ie vanilla) scenario, at each time slot, the virtualized platform is allocated on a different set of processing cores by the scheduler that may not favor IO affinity and implies reestablishing the cache. We elaborate this analysis further in Section~\ref{sec:cross} with respect to other application types as well.


\end{enumerate}

\subsubsection{NoSQL Workload using Apache Cassandra}
Apache Cassandra \cite{abramova2013nosql} is a distributed NoSQL database extensively used to handle Big Data in the cloud. We evaluate it in this study as an application type that demands compute, memory, and IO. We configured Cassandra exclusively on one execution platform and used its native stress tool, \emph{Cassandra-stress} \cite{shirinbab2017performance}\cite{abramova2013nosql}, to submit 1,000 synthesized database operations within one second. A set of 100 threads, each one simulating one user, were spawned by Cassandra-stress. To make the imposed overhead stand out, we put Cassandra under extreme pressure by forcing a quarter of the synthesized requests as the \emph{write} operations and the rest as the \emph{read} operations. Then, we calculated the average execution time (a.k.a response time) of all the synthesized operations. To capture the randomness in the results, caused by the execution platforms, we conducted the experiment 20 times for each instance type and the mean and 95\% confidence interval of the results are reported.

\begin{figure}[b]
 \vspace{-7px}
  \centering
  \includegraphics[width=0.45\textwidth]{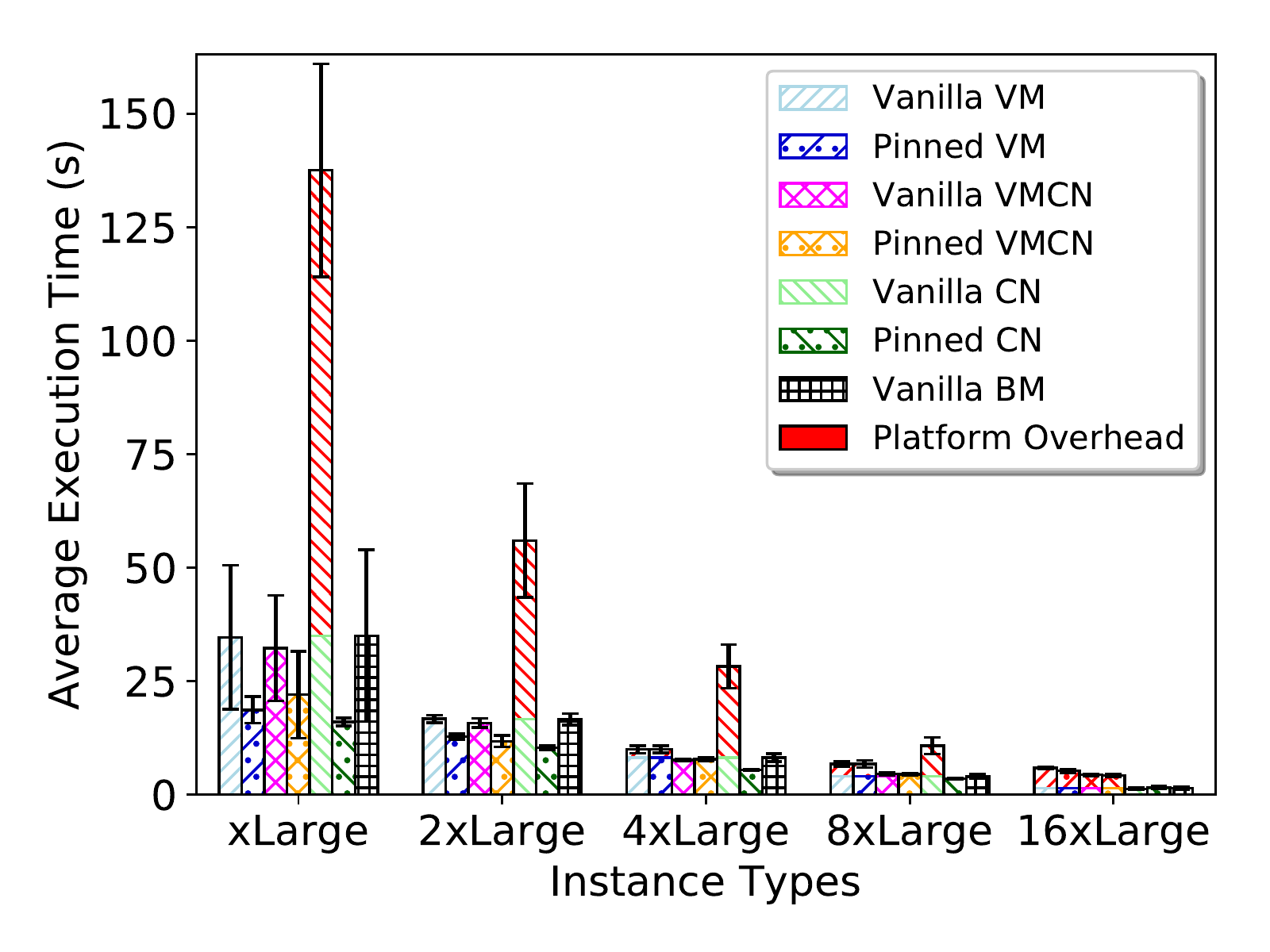}
  \caption{\small{Comparing mean execution time (aka response time) of Cassandra workload (in seconds) on different execution platforms. Horizontal axis represents the number of CPU cores in the form of different instance types. Note that the execution time for the \texttt{Large} instance type is out of range and unchartable. }
  \label{Fig:Cassandra-all}}
\end{figure}

Results of this evaluation are shown in Figure \ref{Fig:Cassandra-all}. Note that, for the \texttt{Large} instance type, the system is overloaded and thrashed and the results are out of range. As such, to be able to concentrate on the imposed overhead of other instance types, we excluded the results of the \texttt{Large} type. Our observations and analysis for this experiment are as follows: 

\begin{enumerate}[label=\roman*.]
\item Vanilla CN imposes the largest overhead---3.5 times or more with respect to BM. This overhead is even higher than the similar phenomenon we observed for WordPress (in Figure~\ref{Fig:Wordpress-all}). The reason for this behavior is the higher volume of IO operations in Cassandra rather than WordPress. However, the CN overhead diminishes for instances with larger number of cores. As we had similar observations for other application types, we defer further analysis of this phenomenon to Section~\ref{sec:cross}. 

\item In contrast to vanilla CN, pinned CN imposes the lowest overhead. This confirms our earlier observations on the positive impact of pinning on IO intensive applications. Surprisingly, we see that for \texttt{$\times$Large}---\texttt{4$\times$Large} instance types, pinned CN can even beat BM. The reason is the BM scheduler is oblivious to IO affinity and the extreme volume of IO operations makes BM perform worse than the pinned CN. For the same reason, we can see that offering a lower execution time is not limited to only pinned CN, but it is also noticed in any pinned virtualized platform.

\item By increasing the number of cores, the impact of pinning is diminished to the extent that for \texttt{8$\times$Large} and \texttt{16$\times$Large} instance types, there is no improvement between the pinned virtualized platforms and their vanilla counterparts. We believe this is because in virtualized platforms with larger instance types: (A) there are fewer scheduling options within the host machine, hence, the scheduling overhead is mitigated; (B) cache and IO affinity are improved, because in each scheduling time slot, it is likely that a virtualized platform is assigned to the same set of processing cores.

\item For all VM-based platforms whose instance type is \texttt{8$\times$Large} and beyond, there is an increased overhead with respect to BM. As noted in the previous point, for larger instances, the overhead of IO diminishes, hence, the execution time is dominated by the CPU processing time. This makes the situation similar to CPU intensive application types (\eg FFmpeg) where VM-based platforms impose a considerable overhead. 

\end{enumerate}

\section{Cross-Application Overhead Analysis}\label{sec:cross}

Building upon application-specific observations conducted in the previous section, in this part, we further analyze the root causes of the imposed overhead by various virtualization platforms across different application types. We also carry out additional experiments to verify our findings.

By comparing the results and studying overhead behavior across all application types and execution platforms, the following two categories of the imposed overhead can be distinguished:
\paragraph*{1) \textbf{Platform-Type Overhead (PTO)}}
This kind of overhead pertains to the type of virtualized platform and its ratio remains constant, irrespective of the instance type it is deployed on. For instance, in both FFmpeg (Figure~\ref{Fig:FFmpeg-all}) and Cassandra (for instance types greater than \texttt{4$\times$Large} in Figure~\ref{Fig:Cassandra-all}), the overhead ratio of VM remains the same across various instance types. This type of overhead is caused by the abstraction layers of the virtualized platforms and are reflected when CPU intensive applications (\eg FFmpeg) are deployed. As the source of this type of overhead pertains to virutalization layers, pinning cannot mitigate it drastically. This type of overhead is a known issue and has been studied in prior works \cite{felter2015updated,ccpe14amini,spe14amini}.

\paragraph*{2) \textbf{Platform-Size Overhead (PSO)}}
This type of overhead is diminished by increasing the number of cores assigned and is specific to vanilla CN platform. PSO is observed throughout all the studied application types, particularly, when a container is assigned a small portion of the host CPU cores. Our hypothesis is that the ratio of the container cores to the host cores is a decisive factor on the magnitude of the imposed overhead. This behavior of containers has also been identified and reported for Docker by IBM~\cite{IBM}. In particular, for IO-bound applications, the overhead even exceeds VM platforms. As an instance, for \texttt{Large} instance type, the overhead ratio of vanilla CN to vanilla VM for WordPress application is 2.4 (see Figure~\ref{Fig:Wordpress-all}) and it is 3.7 for Cassandra (see Figure~\ref{Fig:Cassandra-all}). Importantly, pinning can considerably mitigate this type of overhead. Since this type of overhead has not been investigated before, we elaborate on its root causes in the next subsections. 

\subsection{The Impact of Container-to-Host Core Ratio (CHR) on PSO}
To analyze the impact of container size on PSO, for a given container, we define \emph{Container-to-Host Core Ratio} (CHR) as the ratio of its assigned cores to the total number of host cores. To evaluate the impact of CHR, we choose FFmpeg as the application type, because it does not impose additional IO overhead and our analysis is concentrated on PSO.
We configure a CN platform of \texttt{4$\times$Large} type on two homogeneous hosts, with 16 and 112 cores, respectively. Then, we measure the mean execution time of the FFmpeg workload (described in Section~\ref{sec:PerfStudyanalysis}) on these configurations. 

\begin{figure}[b]

  \centering
  \includegraphics[width=0.45\textwidth]{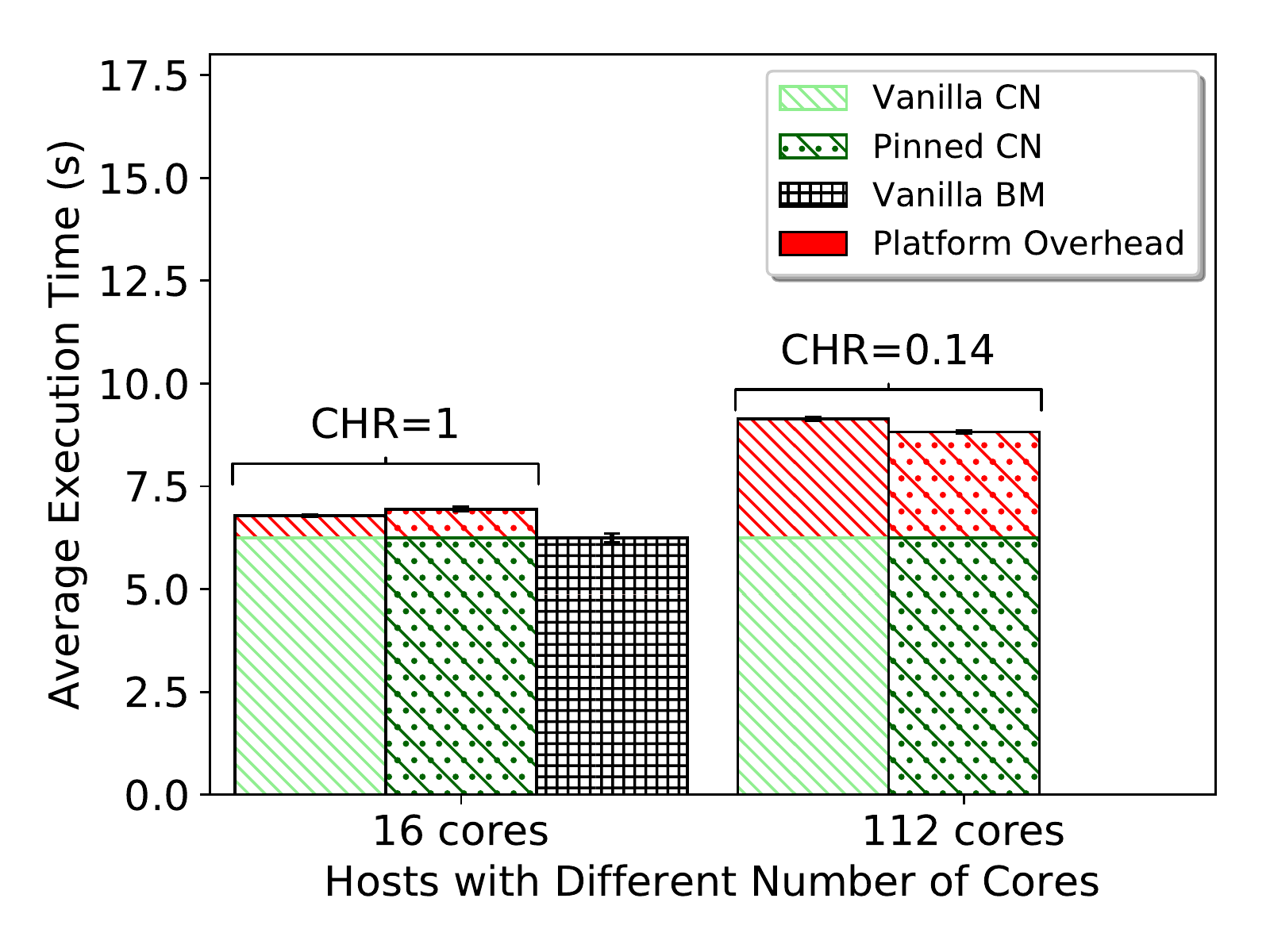}
  \caption{\small{Evaluating the impact of CHR on the overhead of a vanilla and a pinned CN platform on two homogeneous hosts with 16 and 112 cores. The vertical axis shows the mean execution time (in seconds) and the horizontal axis shows the host's number of cores.}}
  \label{Fig:16coreBM}
      \vspace{-7px}
\end{figure}

Results of this experiment are shown in Figure \ref{Fig:16coreBM}. 
The first set of bars represent CHR=1 and the second set represent CHR=0.14. In addition to the CN platform, we report the result for the BM platform with 16 cores. This enables us to study the overhead of a CN platform that is as big as the host machine.

In this figure, we observe that although the CN platforms have the same number of cores, on the larger host (with 112 cores) they impose a higher overhead in compare with the case that they are deployed on the smaller host. We can conclude that a container with a lower value of CHR imposes a larger overhead (PSO) to the system. As illustrated in the figure, pinning does not imply any statistically significant difference in the imposed overhead.

Throughout Section~\ref{sec:PerfStudyanalysis}, we repeatedly observed  the impact of increasing CHR on mitigating PSO across different application types. That is, for larger CN platforms, the execution time approaches BM, even more rapidly than other virtualized platforms. 

The question arises in this case is that, for a given container that processes a certain application type, how to know the suitable value of CHR? In this work, we address this question based on our observations for different application types. However, we note that answering this question theoretically is an interesting future research work. 
In WordPress, the PSO starts to vanish when the CN is configured in the range of [\texttt{4$\times$Large} , \texttt{8$\times$Large}] that implies $0.14 < CHR < 0.28$. A similar observation was made in the IBM report for their proprietary IBM WebSphere web server \cite{IBM}. Similar analysis for FFmpeg and Cassandra indicate suitable CHR should be in the range of $0.07 < CHR < 0.14$ and $0.28 < CHR < 0.57$, respectively. From this analysis, we conclude that IO intensive applications require a higher CHR value than the CPU intensive ones.

The estimated ranges of CHR can be used by cloud administrators to configure containers such that their imposed overhead is minimized. This can potentially benefit both cloud users (by reducing their incurred cost and response time) and cloud providers (by lowering their energy consumption).

\subsection{The Impact of Container Resource Usage Tracking on PSO}
Although CHR explains the diminishing PSO for containers with larger instances, it does not explain the high PSO of small vanilla CNs (\ie those with low CHR) that we observed for all task types. Recall (from Figures \ref{Fig:FFmpeg-all}, \ref{Fig:Wordpress-all}, and \ref{Fig:Cassandra-all}) that pinning remarkably mitigate the PSO for low-CHR CNs. Accordingly, our hypothesis is that the high PSO is attributed to CPU provisioning model. 

We observed in our experiments that, for small vanilla CN (those with 2 cores), \os~scheduler allocates all available CPU cores of the host machine (112 cores) to the CN process. In this circumstance, cgroups has to assure that the cumulative CPU usage of the process does not exceed its designated quota. This means that in each scheduling event, vanilla CN undergoes the overhead of both OS scheduling (that implies process migration) and cgroups (for resource usage tracking). We realized that cgroups is an atomic (kernel space) process \cite{Kernel.org-cgroup}\cite{Josef} and each invocation of it implies one transition from the user-mode to the kernel-mode, which incurs a considerable overhead. Furthermore, we noticed that, in amortizing the process of a small container across all available CPU cores, the process footprint on each core is a tiny fraction of the whole process. Specifically, for small containers, we observed that the overhead of cgroups tasks reaches to the point that it dominates the container process. In particular, because cgroups is an atomic process, the container has to be suspended, until tracking and aggregating resource usage of the container is complete. 

In contrast to the vanilla mode, CPU pinning avoids the overhead of host \os~scheduling and cgroups. In the presence of pinning, as the allocated set of processing cores do not change at each scheduling event, there is less demand for cgroups invocation, hence, the imposed overhead is mitigated. 

\subsection{The Impact of IO Operations on PSO}
The comparison of FFmpeg (particularly, \texttt{$\times$Large} in Figure \ref{Fig:FFmpeg-all}) against WordPress and Cassandra (particularly, \texttt{$\times$Large} in Figures \ref{Fig:Wordpress-all} and \ref{Fig:Cassandra-all}) illustrates that, in both WordPress and Cassandra, the PSO of vanilla CN (\ie the overhead part, colored in \textcolor{red}{red}) is drastically higher than FFmpeg. As both WordPress and Cassandra are IO-bound applications, our hypothesis is that the extra overhead pertains to performing the IO operations.

As a CPU-bound application, FFmpeg has a predictable behavior and fully utilizes their CPU quota in each time slot. In contrast, the IO-bound applications often do not make a complete use of their CPU quota, because they are interrupted and switch to the  \emph{pending} state, until they complete their IO operation. For example, in WordPress, each web request triggers at least three Interrupt Requests (IRQs): to read from the network socket; to fetch the requested HTML file from disk; and to write back to the network socket. Each IRQ implies the overheads to accomplish a set of scheduling actions (to enqueue, dequeue, and pick the next task 
) and transitioning to the kernel mode (to perform the IO operation).

Once an interrupt is served, to avoid cache line bouncing and reestablishing IO channels, OS scheduler makes its best effort to resume the interrupted tasks on the same set of cores. However, in the event that the process is assigned to a different set of cores, a significant overhead is imposed to reload L1 and L2 caches and establish new IO channels. As noted in the previous section, vanilla CN instances with smaller CHR are more prone to be allocated on a different set of cores, hence, experiencing even a more significant overhead, when compared with larger instances. 

\subsection{The Impact of Multitasking on PSO}
Analysis of the overhead across different application types, in particular FFmpeg versus WordPress, brings us to the hypothesis that the number of processes increases the imposed PSO. To verify the hypothesis and figure out the importance of this factor, in this part, we conduct an experiment to analyze the impact of number of FFmpeg processes on its PSO. The reason that we compare FFmpeg with itself is to eliminate the impact of differences in application characteristics and, instead, single out the impact of number of processes. 

For this experiment, we examined FFmpeg on \texttt{4$\times$Large} CN instance types to change the codec of the video file used in Section~\ref{subsec:ffmpeg}. We studied two scenarios: (A) the source video file is a 30-second (large) video; and (B) splitting the same source video into 30 video files of the same size (one-second each) and process them in parallel.

\begin{figure}[h]
   \vspace{-3px}
  \centering
  \includegraphics[width=0.45\textwidth]{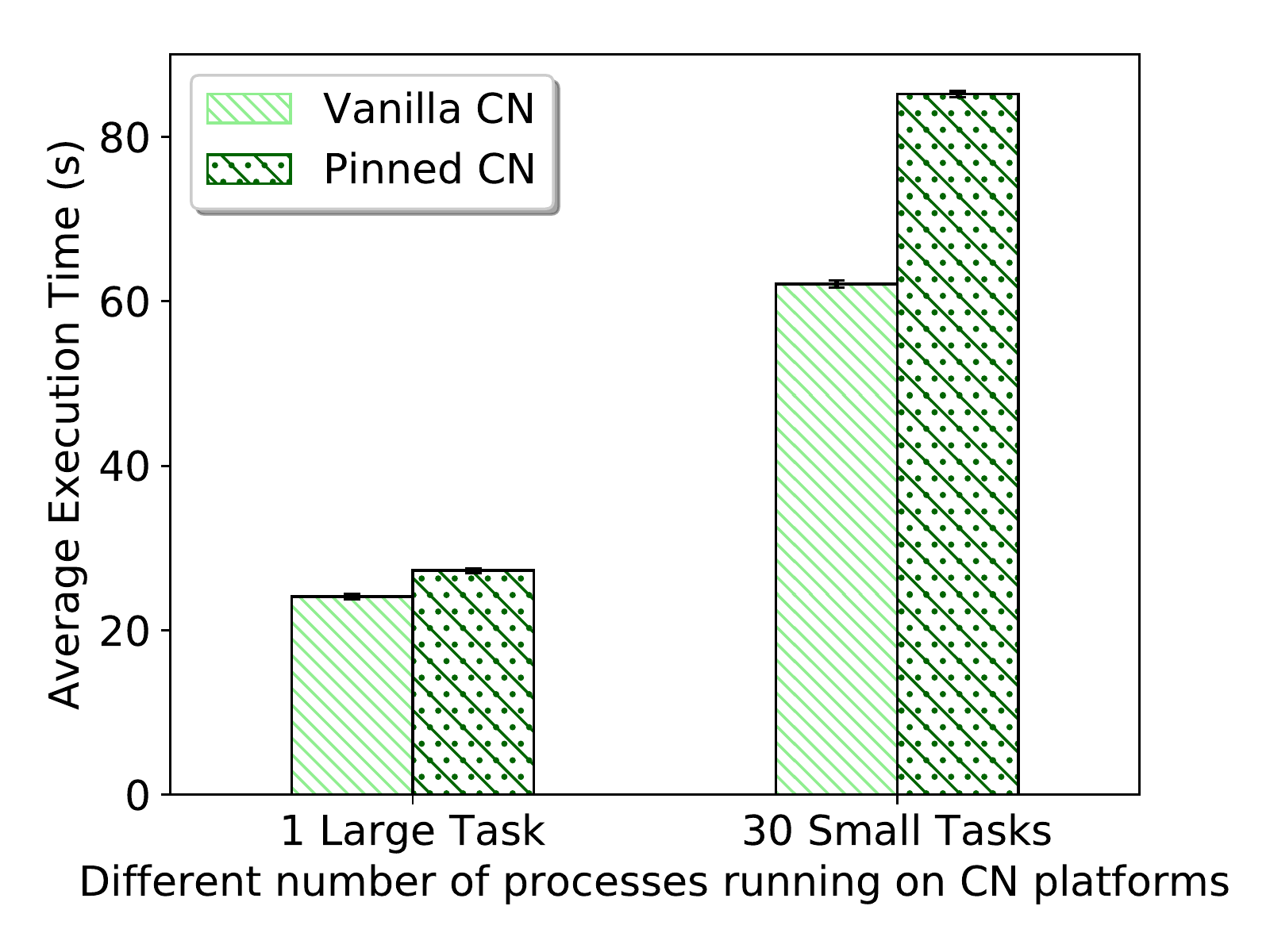}
  \caption{\small{Comparing the impact of number of processes on the imposed overhead of \texttt{4$\times$Large} CN instance. The vertical axis shows the mean executing times (in Seconds) and the horizontal axis shows processing of a source video file in two cases: one large video versus partitioning it into 30 small videos.}}
  \label{Fig:LargeSmall}
  \vspace{-5px}
\end{figure}

Comparing the results in Figure~\ref{Fig:LargeSmall} approves our hypothesis and shows that the number of processes results in increasing the PSO of CN platforms. This is because, a higher degree of multitasking increases the overhead imposed by OS scheduler and cgroups to collect resource usage of CNs.

\section{Related Work}\label{sec:relwk}
Container technology has gained a major attention over the past few years, thus, several research works have been undertaken to evaluate their performance in contrast to VMs. Xavier \etal~\cite{xavier2013performance} compared the Linux VServer, OpenVZ, and Linux Containers (LXC) with Xen hypervisor using synthetic workloads such as LINPACK, STREAM and IOzone. They observed that containers can provide a near-native performance for the mentioned workloads. However, this work does not consider popular cloud applications and platforms. In another study, Shirinbab \etal~\cite{shirinbab2017performance} compare the performance of Docker, VMware ESXi, and Bare-Metal. They evaluated the performance of these platforms for Apache Cassandra database. Their work shows that in terms of disk I/O, containers incur a performance penalty, specially for disk write operations.

In \cite{saha2018evaluation}, authors evaluated the network overhead of Docker and Singularity containers using HPCG and miniFE workloads. Rudyy \etal~\cite{rudyy2019containers} also compared three different container technologies on computational fluid dynamics (CFD) and concluded Singularity is the suitable container solution for HPC workloads that provide the same execution time as Bare-Metal. Mavridis and Karatza~\cite{mavridis2017performance} studied containers on VMs configuration. Similar to \cite{xavier2013performance}, they evaluated Docker on top of KVM VMs using well-known benchmarking tools and explored the overhead in comparison with Bare-Metal. They concluded adding another virtualization layer that helps for easier management and system maintenance, but incurs a performance cost.

Several research works compared containers versus VMs and Bare-Metal \cite{xavier2013performance,shirinbab2017performance,saha2018evaluation,rudyy2019containers,ZhengLi, Morabito,felter2015updated,Barik}. Only few of these studies explored VMCN configuration, which is a popular platform in the cloud \cite{mavridis2017performance}. In addition, majority of the comparisons are performed using synthetic workloads and benchmarking tools such as LINPACK, STREAM, and NetPerf that exhibit different  behavior than real cloud-based workloads.

\section{Summary and Best Practices}\label{sec:conclsn}

In this study, the performance overhead imposed by different virtualized platforms commonly used in the cloud was studied. Four popular application types, namely video processing and parallel processing (MPI), web processing, and No-SQL were studied. 
The study revealed that: \textbf{(A)} application characteristic (\ie IO-bound versus CPU-bound) is decisive on the imposed overhead of different virtualization platforms. \textbf{(B)} CPU pinning can reduce the overhead of virtualized platforms, particularly for IO-bound applications running on containers. \textbf{(C)} The proportion of container cores to the host cores (we named it CHR) plays a significant role on the overhead of containers. Greater CHR values reduce the overhead. \textbf{(D)} Although containers are known to impose a lower overhead than VMs, the opposite was observed for containers with very low CHRs. \textbf{(E)} Containers on top of VMs (called VMCN) impose a lower overhead for IO intensive applications. \textbf{(F)} Increasing the level of multitasking amplifies the overhead of containers.

In addition, this study provides the following set of \textbf{Best Practices} that can help cloud architects to efficiently configure cloud systems based on their application types:
\begin{shaded*}
\begin{enumerate}
\item Avoid instantiating small vanilla containers (with one or two cores) for any type of application.\vspace{3 mm}
\item For CPU intensive applications (\eg FFmpeg), pinned containers impose the least overhead. \vspace{3 mm}
\item If VMs are being utilized for CPU-bound applications, do not bother pinning them. It neither improves the performance, nor decreases the incurred cost.\vspace{3 mm}
\item For IO intensive applications, if pinned container is not a viable option, then use container within VM (VMCN). It imposes a lower overhead than a VM or a vanilla container.\vspace{3 mm}
\item To minimize the overhead of containers, for CPU intensive applications configure them with $0.07<CHR<0.14$ and for IO intensive applications use $0.14<CHR<0.28$. If the application is ultra IO intensive (\eg Cassandra), even a higher CHR ($0.28<CHR <0.57$) is suggested.

\end{enumerate}
\end{shaded*}
In the further, we plan to extend the study to incorporate the impact of network overhead. In addition, we plan to provide a mathematical model to measure the overhead of a given virtualization platforms based on the isolation level it offers.

\section*{Acknowledgment}
\small{We thank reviewers of the manuscript. The research was supported
by the Louisiana Board of Regents under grant number LEQSF (2017-
20)-RD-B-06, Perceptive Intelligence LLC, and Amazon Cloud (AWS) research credit.}

%
\linespread{0.95} 
 \bibliographystyle{IEEEtran} 
 \bibliography{paper}


\end{document}